\documentclass[sigconf]{acmart}
\usepackage{diagbox}
\usepackage{enumitem}
\AtBeginDocument{%
  \providecommand\BibTeX{{%
    \normalfont B\kern-0.5em{\scshape i\kern-0.25em b}\kern-0.8em\TeX}}}

\copyrightyear{2021}
\acmYear{2021}
\setcopyright{acmcopyright}\acmConference[ICAIF'21]{A2nd ACM International Conference on AI in Finance}{November 3--5, 2021}{Virtual Event, USA}
\acmBooktitle{2nd ACM International Conference on AI in Finance (ICAIF'21), November 3--5, 2021, Virtual Event, USA}
\acmPrice{15.00}
\acmDOI{10.1145/3490354.3494415}
\acmISBN{978-1-4503-9148-1/21/11}

\usepackage{algorithmic}
\usepackage{algorithm}
\usepackage{bm}

\begin{document}

% xFinRL: 

\title{Explainable Deep Reinforcement Learning for Portfolio Management: An Empirical Approach}

\author{Mao Guan}
\email{mg3844@columbia.edu}
\affiliation{%
  \institution{ Computer Science, Columbia University}
  \city{New York City}
  \state{New York}
}

\author{Xiao-Yang Liu}
\authornote{Equal contribution.}
\authornote{Corresponding author.}
\email{xl2427@columbia.edu}
\affiliation{%
  \institution{Electrical Engineering, Columbia University}
  \city{New York City}
  \state{New York}
}

%\title{Deep Ensemble Reinforcement Learning for Automated Stock Trading}

%\author{Mao Guan}
%\email{mao.guan@columbia.edu}
%\affiliation{%
%  \institution{ Columbia University}
%  \city{New York City}
%  \state{New York}
%}

%\author{Xiao-Yang Liu}
%\authornote{Co-primary author with equal contribution.}
%\authornote{Corresponding author.}
%\email{xl2427@columbia.edu}
%\affiliation{%
%  \institution{Electrical Engineering, Columbia University}
%  \city{New York City}
%  \state{New York}
%}


% \newcommand{\yanglet}[1]{\textcolor{red}{[Yanglet: #1]}}
% \newcommand{\brucey}[1]{\textcolor{red}{[BruceYang: #1]}}

%% 我的贡献点是什么？挑战是什么？
%% 先sale我的课题，我们连做这件事情为什么这么重要？关键词是什么？不是只关心盈利，没有上线，为什么有机会盈利？为什么是个promising的approach，stock market的dynamic的，stock market他们是random walk, stock trading dynamic decision making在stochastic process上的，挑战是为什么挑战？股票是很多人参与的一个群体行为，很难预测，你只预测股价没有用，还要预测其他人的play
%% The abstract is a short summary of the work to be presented in the article.
%% 用多种历史数据，首先是建议一个准确的stock market env，trade一个trading agent，用的是DRL的方法，仿真平台
%%第一个是仿真环境：模拟环境和回测
%金融的难度:点到底点到哪里
%环境构建，和方法构建
%为金融的解决问题是什么：赚钱是最终目的，
%更有意义：把baseline都调好，深度增强学习是个黑盒子，增加解释性
%把算法的观察角度拿出来
%怎么来评估，深度学习方法起作用，然后达到很好的效果呢？benchmark，大盘，optimal比一比
%单只股票上调优，到多只股票，目的是解决，黑盒子跑的好和跑的坏，都不知道哪里出问题，但只股票是为了创建参照物和可比性，在这个基础上去调试算法，为了礼节性和调优
%环境，黑盒，回测，baseline

\begin{abstract}

Deep reinforcement learning (DRL) has been widely studied in the portfolio management task. However,  it is challenging to understand a DRL-based trading strategy because of the black-box nature of deep neural networks. In this paper, we propose an empirical approach to explain the strategies of DRL agents for the portfolio management task.
First,  we use a \textit{linear model in hindsight} as the \textit{reference model}, which finds the best portfolio weights by assuming knowing actual stock returns \textit{in foresight}. In particular, we use the coefficients of a linear model in hindsight as the \textit{reference feature weights}. Secondly, for DRL agents, we use integrated gradients to define the feature weights, which are the coefficients between reward and features under a linear regression model. Thirdly, we study the prediction power  in two cases, single-step prediction and multi-step prediction. In particular, we quantify the prediction power by calculating the linear correlations between the feature weights of a DRL agent and the \textit{reference feature weights}, and similarly for machine learning methods. Finally, we evaluate a portfolio management task on Dow Jones 30 constituent stocks during 01/01/2009 to 09/01/2021. Our approach empirically reveals that a DRL agent exhibits a stronger multi-step prediction power than machine learning methods.

\end{abstract}




\maketitle

%%%%%%%%%%%%%%%%%%%%%%%%%%%%%%%%%%%%%%%%%%%%%%%%%
\section{Introduction}

The explanation \cite{jaeger2020understanding} of a portfolio management strategy is important to investment banks, asset management companies and hedge funds. It helps traders understand the potential risk of a certain strategy. However, it is challenging to explain a DRL-based portfolio management strategy due to the black-box nature of deep neural networks.

Existing DRL-based portfolio management works focus on enhancing the performance.  A typical DRL approach of portfolio management consists of three steps as described in \cite{liu2020finrl, liu2021finrl, liu2021neofinrl, gpu_podracer_nips_2021, finrl_podracer_2021}. First, select a  pool of possibly risky assets.  Secondly, specify the state space, action space and reward function of the DRL agent. Finally, train a DRL agent to learn a portfolio management strategy. Such a practical approach, however, does not provide explanation to the portfolio management strategy.

In recent years, explainable deep reinforcement learning methods have been widely studied. Quantifying how much a change in input would influence the output is important to understand what contributes to the decision-making processes of the DRL agents. Thus, saliency maps \cite{tjoa2020survey} are adopted to provide explanation. However, these approaches are mainly available in computer vision, natural language processing and games \cite{atrey2019exploratory, heuillet2021explainability, madumal2020explainable}. They have not been widely applied in financial applications yet.
Some researchers  \cite{cong2021alphaportfolio} explain the DRL based portfolio management strategy  using an attention model. However, it does not explain the decision-making process of a DRL agent in a proper financial context.

In this paper, we take an empirical approach to explain the portfolio management strategy of DRL agents. Our contributions are summarized as follows
\begin{itemize}[leftmargin=*]
   \item We propose a novel empirical approach to understand the strategies of DRL agents for the portfolio management task. In particular, we use the coefficients of a \textit{linear model in hindsight} as the \textit{reference feature weights}.
   \item For a deep reinforcement learning strategy, we use integrated gradients to define the feature weights, which are the coefficients between the reward and features under a linear regression model.
   \item We quantify the prediction power by calculating the linear correlations between the feature weights of a DRL agent and the reference feature weights, and similarly for conventional machine learning methods. Moreover, we consider both the single-step case and multiple-step case.
   %\item We measure the DRL agents' and machine learning methods' prediction power by comparing with reference model using correlation coefficients. 
    \item We evaluate our approach on a portfolio management task with Dow Jones 30 constituent stocks during 01/01/2009 to 09/01/2021. Our approach empirically explains that a DRL agent achieves better trading performance because of its stronger multi-step prediction power.
\end{itemize}

The remainder of this paper is organized as follows. In Section 2, we review existing works on the explainable deep reinforcement learning. In Section 3, we describe the problem formulation of a DRL-based portfolio management task. In Section 4, we present the proposed explanation method. In Section 5, we show quantitative experimental results of our empirical approach. Finally, the conclusion and future work are given in Section 6.

\section{Related Works}

Gradient based explanation methods are widely adopted in the saliency maps \cite{tjoa2020survey}, which quantify how much a change in input would influence the output. We review the related works of gradient based explanation for deep reinforcement learning. 
\begin{itemize}[leftmargin=*]
    \item \textbf{Gradient} $\odot$  \textbf{Input} \cite{shrikumar2017learning} is the element-wise product of the gradient and the input. It provides explanation by visualizing the product as heatmap.
    \item \textbf{Integrated Gradient (IG)} \cite{pmlr-v70-sundararajan17a}.
    It integrates the gradient of the output with respect to input features. For an input $\bm{x} \in \mathbb{R}^n$, the $i$-th entry of integrated gradient is defined as  
    \begin{equation}\label{eq:IG_def}
    \text{IG}(\bm{x})_{i} \triangleq (\bm{x}_{i} - \bm{x}^{\prime}_{i}) \times \int_{z=0}^{1}\frac{\partial F(\bm{x}^{\prime} + z\cdot(\bm{x} - \bm{x}^{\prime}))}{\partial \bm{x}_{i}}dz,
    \end{equation}
    where $F(\cdot)$ denotes a DRL model, $\bm{x}^{\prime}$ is a perturbed version of $\bm{x}$, say replacing all entries with zeros. It explains the relationship between a model's predictions in terms of its features.
    \item \textbf{Guided Backpropagation (GBP)}
    computes the gradient of the target output with respect to the input
    \cite{springenberg2014striving, zeiler2014visualizing}, and it treats negative gradients as zeros. It provides explanation by visualizing the gradients.
    \item \textbf{Guided GradCAM} \cite{selvaraju2016grad}. 
    It uses the class-specific gradient and the final layer of a convolutional neural network to produce a coarse localization map of the important regions in an image. It provides explanation using a gradient-weighted map.
    \item \textbf{SmoothGrad (SG)} \cite{smilkov2017smoothgrad, pmlr-v70-sundararajan17a}. 
    It creates noisy copies of an input image then averages gradients with respect to these copies. It provides explanation using visual map to identify pixels that strongly influence the final result.
\end{itemize}

Although these gradient based explanation methods are popular, they have not been directly applicable to the portfolio management task yet. Other researchers  \cite{cong2021alphaportfolio} explain the DRL based portfolio management  using an attention model. However, it does not explain the decision-making process of DRL agent in a proper financial context.

% The gradient-based metrics are widely used in important financial concepts such as risk premium \cite{ross2013arbitrage}. We believe that gradient explanation methods may have great potential in financial applications using deep reinforcement learning. William {\em et  al\adddot} \cite{cong2021alphaportfolio} rely on attention model to explain the portfolio management using reinforcement learning.

\section{Portfolio Management Using Deep Reinforcement Learning}

\begin{figure}[t]
\centering
\includegraphics[scale=0.6]{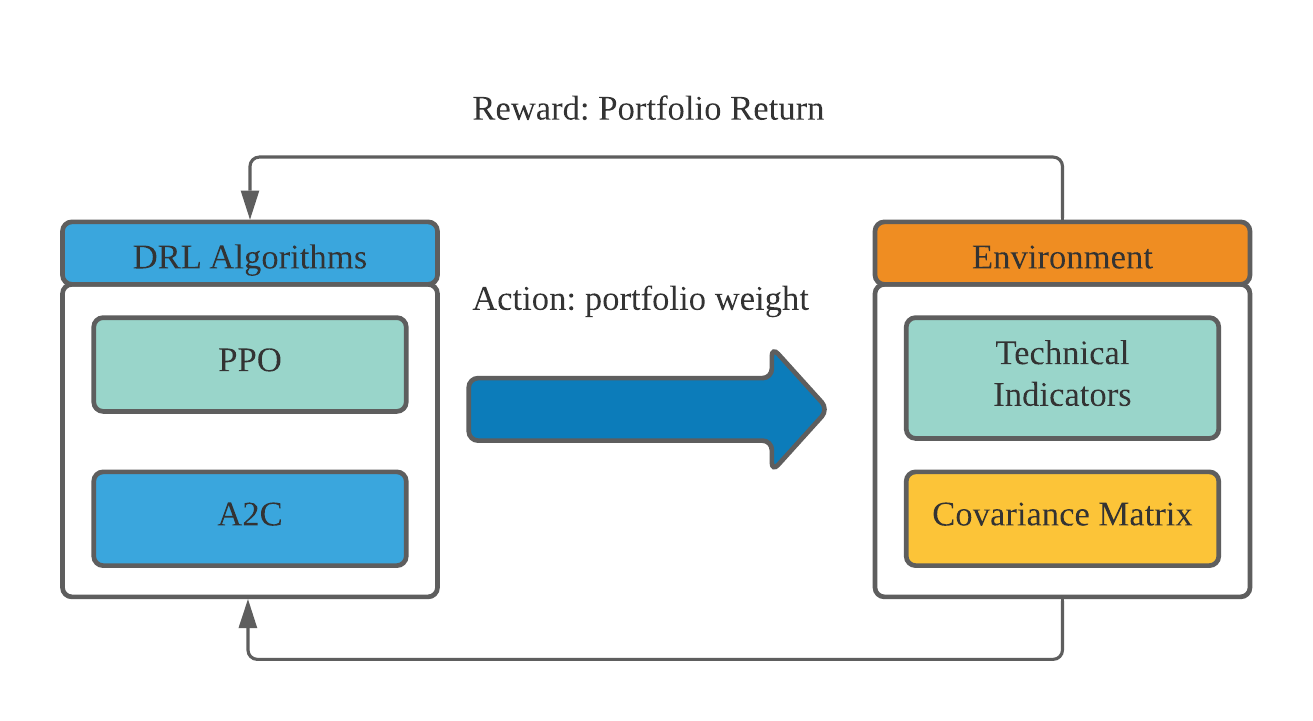}
\caption{Overview of a portfolio management task that uses
deep reinforcement learning.}
\label{overview_pm_drl}
\end{figure}

We first describe a portfolio management task using a DRL agent. Then we define the feature weights using integrated gradients.

\subsection{Portfolio Management Task}

Consider a portfolio with $N$ risky assets over $T$ time slots, the portfolio management task aims to maximize profit and minimize risk. Let $\bm{p}(t) \in \mathbb{R}^{N}$ denotes the closing prices of all assets at time slot $t = 1,..., T$. \footnote{For continuous markets, the closing prices at time slot $t$ is also the opening prices for time slot $t+1$.}The \textit{price relative vector} $\bm{y}(t) \in \mathbb{R}^{N}$ is defined as the element-wise division of $\bm{p}(t)$ by $\bm{p}(t-1)$:
\begin{equation}
    \bm{y}(t) \triangleq \left[ \frac{\bm{p}_{1}(t)}{\bm{p}_{1} (t-1)}, \frac{\bm{p}_{2}(t)}{\bm{p}_{2}(t-1)}, ..., \frac{\bm{p}_{N}(t)}{\bm{p}_{N}(t-1)} \right]^{\top},~~ t =1, .... T,
\label{eq:return_vector}
\end{equation}
where $\bm{p}(0) \in \mathbb{R}^{N}$ is the vector of opening prices at $t = 1$.

 Let $\bm{w}(t) \in \mathbb{R}^{N}$ denotes the portfolio weights, which is updated at the beginning of time slot $t$. Let $v(t) \in \mathbb{R}$ denotes the portfolio value at the beginning of time slot $t+1$. \footnote{Similarly $v(t)$ is also the portfolio value at the ending of time slot $t$.}
Ignoring the transaction cost, we have the \textit{relative portfolio value} as the ratio between the portfolio value at the ending of time slot $t$ and that at the beginning of time slot $t$,
\begin{equation}
    \frac{v(t)}{v(t-1)} = \bm{w}(t)^{\top} \bm{y}(t),
\end{equation}
where $v(0)$ is the initial capital. The \textit{rate of portfolio return} is 
\begin{equation}
    \rho(t) \triangleq \frac{v(t)}{v(t-1)} -1 = \bm{w}(t)^{\top} \bm{y}(t) - 1,
\end{equation}
while correspondingly the \textit{logarithmic rate of portfolio return} is 
\begin{equation}
    r(t) \triangleq \ln \frac{v(t)}{v(t-1)} = \ln(\bm{w}(t)^{\top}\bm{y}(t)).
\label{eq:reward}
\end{equation}

The risk of a portfolio is defined as the variance of the rate of portfolio return $\rho(t)$:
\begin{equation}
\begin{split}
        &\text{Risk}(t)  \triangleq \text{Var}(\rho(t)) = \text{Var}(\bm{w}(t) ^{\top}\bm{y}(t) - 1) \\
        &= \text{Var}(\bm{w}(t) ^{\top}\bm{y}(t)) =\bm{w}(t) ^{\top}~\text{Cov}(\bm{y}(t))~\bm{w}(t)\\
            &=\bm{w}(t)^{\top}~\bm{\Sigma}(t)~\bm{w}(t),
\end{split}
\end{equation}
where  $\bm{\Sigma}(t) = \text{Cov}(\bm{y}(t)) \in  \mathbb{R}^{N \times N}$ is the covariance matrix of the stock returns at the end of time slot $t$. 
If there is no transaction cost, the final portfolio value is
\begin{equation}
    v(T) = v(0)~\exp\left( \sum\limits_{t=1}^{T} r(t) \right) = v(0)~ \prod\limits_{t=1}^{T} \bm{w}(t)^{\top}\bm{y}(t).
\label{eq:portfolio_value}
\end{equation}

% In real world, regression models\cite{ma2021portfolio,yu2020portfolio} are used to predict the stocks returns with financial factors\cite{feng2017taming}.

% Based on Capital Asset Pricing Model (CAPM) \cite{fama2004capital}, financial factors \cite{feng2017taming} are treated as features to predict stock returns with regression models, say $\widehat{\bm{y}}(t) \in \mathbb{R}^{N}$ is an estimate of the return vector, $\bm{y}(t)$ in (\ref{eq:return_vector}). 

The portfolio management task \cite{boyd2017multi, enwiki:1043516653} aims to find a portfolio weight vector $\bm{w}^{*}(t) \in \mathbb{R}^{N}$ such that

\begin{equation}\label{eq:opt_problem0}
\begin{split}
\bm{w}^{*}(t) \triangleq & \text{argmax}_{\bm{w}(t)}~~~~\bm{w}^{\top}(t) ~ \bm{y}(t) - \lambda ~ \bm{w}^{\top}(t)~ \bm{{\Sigma}}(t) ~ \bm{w}(t),\\
     & \text{s.t.}~~~ \sum_{i=1}^{N} \bm{w}_{i}(t) = 1,~~~~\bm{w}_{i}(t) \in [0, 1],~~~~~~t = 1,...,T
\end{split}
\end{equation}
where $\lambda > 0$ is the risk aversion parameter. Since 
$\bm{y}(t)$ and $\bm{\Sigma}(t)$ are revealed at the end of time slot $t$. We estimate them at the the beginning of time slot $t$.  

We use $\widehat{\bm{y}}(t) \in \mathbb{R}^{N}$ to estimate  the price relative vector $\bm{y}(t)$ in (\ref{eq:opt_problem0}) by applying a regression model on predictive financial features \cite{feng2017taming} based on  Capital Asset Pricing Model (CAPM) \cite{fama2004capital}. 
We use $\widehat{\bm{\Sigma}}(t)$, the sample covariance matrix, to  estimate covariance matrix $\bm{\Sigma}(t)$ in (\ref{eq:opt_problem0}) using historical data.

Then, at the beginning of time slot $t$, our goal is to find  optimal portfolio weights 
\begin{equation}\label{eq:opt_problem}
\begin{split}
\bm{w}^{*}(t) \triangleq & \text{argmax}_{\bm{w}(t)}~~~~\bm{w}^{\top}(t) ~ \widehat{\bm{y}}(t) - \lambda ~ \bm{w}^{\top}(t)~ \widehat{\bm{{\Sigma}}}(t) ~ \bm{w}(t),\\
     &\text{s.t.}~~~ \sum_{i=1}^{N} \bm{w}_{i}(t) = 1,~~~~\bm{w}_{i}(t) \in [0, 1],~~~~~~t = 1,...,T.
\end{split}
\end{equation}
    % \begin{align}
    %     \bm{u}=\begin{bmatrix}
    %         \bm{u}_{1} \\
    %         \bm{u}_{2} \\
    %         \vdots  \\
    %         \bm{u}_{I}
    %     \end{bmatrix}, 
    % \end{align}

% \begin{figure*}[t]
% \includegraphics[width=0.5\textwidth]{figs/integrated_gradient.png}

% \caption{Overview of integrated gradient in DRL.}
% \label{fig:intgrad}
% \end{figure*}
\begin{figure}[t]
\centering
\includegraphics[scale=0.33]{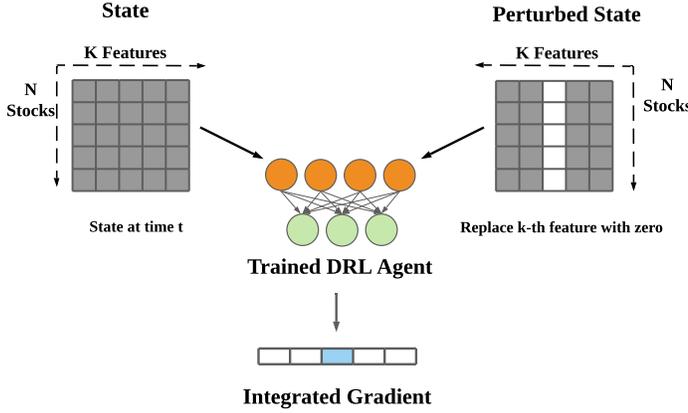}
\caption{Feature weights of a trained DRL agent.}
\label{int_grad}
\end{figure}

\subsection{Deep Reinforcement Learning for Portfolio Management}

We describe how to use deep reinforcement learning algorithms for the portfolio management task, by specifying the state space, action space and reward function. We use a similar setting as in the open-source {FinRL} library \cite{liu2020finrl}\cite{liu2021finrl}.  

\textbf{State space} $\mathcal{S}$ describes an agent's perception of a market.  The state at the beginning of time slot $t$ is
\begin{equation}
    \bm{s}(t) = [\bm{f}^{1}(t),  ... , \bm{f}^{K}(t), \widehat{\bm{\Sigma}}(t)] \in \mathbb{R}^{ N \times (N+K)}, ~~~~~~t = 1,...,T,
\end{equation}
where  $\bm{f}^{k}(t) \in \mathbb{R}^{N}$ denotes the vector for the $k$-th feature at the beginning of time slot $t$.

\textbf{Action space} $\mathcal{A}$ describes the allowed actions an agent can take at a state. In our task, the action $\bm{w}(t) \in \mathbb{R}^{N}$  corresponds to  the portfolio weight vector decided at the beginning of time slot $t$ and should satisfy the constraints in (\ref{eq:opt_problem}). We use a softmax layer as the last layer to meet the constraints.
%It follows the constraints: 1) the sum of weights equal to 1, and 2) range of each weight is between 0 and 1. 

\textbf{Reward function}. The reward function $r(\bm{s}(t),\bm{w}(t),\bm{s}(t+1))$ is the incentive for an agent to learn a profitable policy. We use the logarithmic rate of portfolio return in (\ref{eq:reward}) as the reward,
\begin{equation}
    r(\bm{s}(t),\bm{w}(t),\bm{s}(t+1)) = \ln(\bm{w}^{\top}(t)\cdot\bm{y}(t)).
\end{equation}
The agent takes $\bm{s}(t)$ as input at the beginning of time slot $t$ and output $\bm{w}(t)$ as the portfolio weight vector. 

% The goal is to maximize the summation of the logarithmic rate of portfolio return, e.g., $\sum_{t=1}^{T}r(t)$ in ($\ref{eq:portfolio_value}$).
% The Q-value $Q^{\pi}(\bm{s}(t), \bm{a}(t))$ is the expected cumulative return at state $\bm{s}(t)$ when taking action $\bm{a}(t)$ according to policy $\pi$:
% \begin{equation}
% \begin{split}
% Q^{\pi}(\bm{s}(t), \bm{a}(t))
% &=\mathbb{E}\left[\sum_{l=0}^{\infty}\gamma^{l}r(\bm{s}(t+l),\bm{a}(t+l),\bm{s}(t+l+1))|\bm{s}(t),\bm{a}(t)\right]\\
% \end{split}
% \end{equation}
% where $\gamma \in (0,1]$ is a discount factor.

\textbf{DRL algorithms}. We use two popular deep reinforcement learning algorithms: Advantage Actor Critic (A2C) \cite{mnih2016asynchronous} and  Proximal Policy Optimization (PPO) \cite{schulman2017proximal}.  A2C \cite{mnih2016asynchronous} utilizes an
advantage function to reduce the variance of the policy
gradient. Its objective function is 
\begin{equation}
        \triangledown J_{\theta}(\theta) = \mathbb{E}\left[\sum_{t=1}^{T} \triangledown_{\theta} \log \pi_{\theta}(\bm{w}(t)|\bm{s}(t))A(\bm{s}(t),\bm{w}(t)) \right],
    \end{equation}
    where $\pi_{\theta}(\bm{w}(t)|\bm{s}(t))$ is the policy network parameterized by $\theta$ and  $A(\bm{s}(t),\bm{w}(t))$ is an advantage function defined as follows
    \begin{equation}
    \begin{split}
        &A(\bm{s}(t),\bm{w}(t)) = Q(\bm{s}(t),\bm{w}(t)) - V(\bm{s}(t))\\
        &= r(\bm{s}(t),\bm{w}(t),\bm{s}(t+1)) + \gamma V(\bm{s}(t+1)) - V(\bm{s}(t)),
    \end{split}
    \end{equation}
    where $Q(\bm{s}(t), \bm{w}(t))$ is the expected reward at state $\bm{s}(t)$ when taking action $\bm{w}(t)$, $V(\bm{s}(t))$ is the value function, $\gamma \in (0,1]$ is a discount factor.
    PPO \cite{schulman2017proximal} is used to control the policy gradient update and to ensure that the new policy will be close to the previous one. It uses a surrogate objective function 
    \begin{equation}
   \begin{split}
       &J^{\text{CLIP}}(\theta) =\\
       &\mathbb{E}(t)[\text{min}(R_{t}(\theta) \widehat{A}(\bm{s}(t), \bm{w}(t)),
       \text{clip}(R_{t}(\theta), 1 - \epsilon, 1 + \epsilon)~\widehat{A}(\bm{s}(t),\bm{w}(t))],
   \end{split}
   \end{equation}
where $R_{t}(\theta) \triangleq \frac{\pi_{\theta}(\bm{w}(t)|\bm{s}(t))}{\pi_{\theta_{old}}(\bm{w}(t)|\bm{s}(t))}$ is the probability ratio between new and old policies, $\widehat{A}(\bm{s}(t), \bm{w}(t))$ is the estimated advantage function, and the clip function $\text{clip}(R_{t}(\theta), 1 - \epsilon, 1 + \epsilon)$ truncates the ratio $R_{t}(\theta)$ to be within the range $[1-\epsilon, 1+\epsilon]$.

\subsection{Feature Weights Using Integrated Gradients}
We use the integrated gradients in (\ref{eq:IG_def}) to measure the feature weights \cite{pmlr-v70-sundararajan17a, tomsett2020sanity}.
For a trained DRL agent,  the integrated gradient \cite{pmlr-v70-sundararajan17a} under policy $\pi$ for the $k$-th feature of the $i$-th asset is defined as 
\begin{equation}
\begin{split}
      & IG(\bm{f}^{k}(t))_{i} \\
      & = (\bm{f}^{k}(t)_{i} - \bm{f}^{k^{\prime}}(t)_{i})\\
      & \times  \int_{z = 0}^{1} \frac{\partial Q^{\pi}(\bm{s}_{k}'(t) + z \cdot (\bm{s}(t) - \bm{s}_{k}'(t)), \bm{w}(t))}{\partial \bm{f}^{k}(t)_{i}} d z\\
      & =\bm{f}^{k}(t)_{i} \cdot  \frac{\partial Q^{\pi}(\bm{s}_{k}'(t) + z^{k,i} \cdot (\bm{s}(t) - \bm{s}_{k}'(t)), \bm{w}(t))}{\partial \bm{f}^{k}(t)_{i}} \cdot (1 - 0)\\
      & = \bm{f}^{k}(t)_{i} \\
      & \cdot  \frac{\partial\mathbb{E}\left[\sum_{l=0}^{\infty}\gamma^{l}\cdot r(\bm{s}^{k,i}(t+l),\bm{w}(t+l),\bm{s}^{k,i}(t+l+1))|\bm{s}^{k,i}(t),\bm{w}(t)\right]}{\partial \bm{f}^{k}(t)_{i} }\\
      & = \bm{f}^{k}(t)_{i} \cdot  \sum_{l=0}^{\infty} \gamma^{l}\cdot \frac{\partial\mathbb{E}\left[ \ln(\bm{w}^{\top}(t+l)\cdot\bm{y}(t+l)) |\bm{s}^{k,i}(t),\bm{w}(t)\right]}{\partial \bm{f}^{k}(t)_{i} }\\
      & \approx \bm{f}^{k}(t)_{i} \cdot  \sum_{l=0}^{\infty} \gamma^{l}\cdot \frac{\partial\mathbb{E}\left[ \bm{w}^{\top}(t+l)\cdot\bm{y}(t+l) - 1 |\bm{s}^{k,i}(t),\bm{w}(t)\right]}{ \partial \bm{f}^{k}(t)_{i} }\\
      & = \bm{f}^{k}(t)_{i} \cdot  \sum_{l=0}^{\infty} \gamma^{l}\cdot \frac{\partial\mathbb{E}\left[ \bm{w}^{\top}(t+l)\cdot\bm{y}(t+l)|\bm{s}^{k,i}(t),\bm{w}(t)\right]}{ \partial \bm{f}^{k}(t)_{i} },
\end{split}
\end{equation}

where the first equality holds by definition in (\ref{eq:IG_def}), the second equality holds because of the mean value theorem \cite{enwiki:1036027918}, the third equality holds because
\begin{equation}
    Q^{\pi}(\bm{s}(t), \bm{w}(t))
\triangleq \mathbb{E}\left[\sum_{l=0}^{\infty}\gamma^{l}\cdot r(\bm{s}(t+l),\bm{w}(t+l),\bm{s}(t+l+1))|\bm{s}(t),\bm{w}(t)\right],
\end{equation}
the approximation holds because $\ln(\bm{w}^{\top}(t)\cdot\bm{y}(t)) \approx \bm{w}^{\top}(t)\cdot\bm{y}(t) - 1$ when $\bm{w}^{\top}(t)\cdot\bm{y}(t)$ is close to 1. $\bm{s}_{k}'(t) \in \mathbb{R}^{N \times  (N+K)}$ is a perturbed version of $\bm{s}(t)$ by replacing the $k$-th feature with an all-zero vector.  $\bm{s}^{k,i}(t)$ is a linear combination of original state and perturbed state $\bm{s}^{k,i}(t) \triangleq \bm{s}_{k}'(t) + z^{k,i}  \cdot (\bm{s}(t) - \bm{s}_{k}'(t))$, where $z^{k,i} \in [0,1]$.

%%%%%%%%%%%%%%%%%%%%%%%%%%%%%%%%%%%%%%%%%%%%%%%%%%%%%%%%

\section{Explanation Method}

We propose an empirical approach to explain the portfolio management task that uses a trained DRL agent.

\subsection{Overview of Our Empirical Approach}
Our empirical approach consists of three parts.
\begin{itemize}
    \item First, we study the portfolio management strategy using feature weights, which quantify the relationship between the reward (say, portfolio return) and the input (say, features). In particular, we use the coefficients of \textit{a linear model in hindsight} as the reference feature weights.
   \item Then, for the deep reinforcement learning strategy, we use integrated gradients to define the feature weights, which are the coefficients between reward and features under a linear regression model
   \item Finally, we quantify the prediction power by calculating the linear correlations between the coefficients of a DRL agent and the reference feature weights, and similarly for conventional machine learning methods. Moreover, we consider both the single-step case andmultiple-step case.
\end{itemize}
% \begin{figure}[t]
% \centering
% \includegraphics[scale=0.2]{figs/RLTrade.png}
% \caption{Overview of stock trading with DQN.}
% \label{trading}
% \end{figure}

%%%%%%%%%%%%%%%%%%%%%%%%%%%%%%%%%%%%%%%%%%%%%%%%%%%%%%%%

\subsection{Reference Feature Weights}

For the portfolio management task, we use a \textit{linear model in hindsight} as a reference model. For a \textit{linear model in hindsight}, a demeon would optimize the portfolio  \cite{brown2020portfolio} with actual stock returns and the \textit{actual sample covariance matrix}. It is the upper bound performance that any linear predictive model would have been able to achieve. 

The \textit{portfolio value relative vector} is the element-wise product of weight and price relative vectors, $\bm{q}(t) \triangleq  \bm{w}(t)$ $\odot$ $\bm{y}(t) \in \mathbb{R}^{N}$, where $\bm{w}(t)$ is the optimal portfolio weight. We represent it as a linear regression model as follows
\begin{equation}
    \bm{q}(t) = \beta_0(t)\cdot [1,...,1]^{\top} + \beta_1(t)\cdot\bm{f}^1(t) + ... +  \beta_K(t)\cdot\bm{f}^K(t) + \bm{\epsilon}(t),
\end{equation}
where  $\beta_k(t) \in \mathbb{R}$ is regression coefficient of the $k$-th feature. $\bm{\epsilon}(t) \in \mathbb{R}^{N}$ is the error vector, where the elements are assumed to be independent and normally distributed.
%, $\bm{\epsilon}(t)_{i}  \sim \mathcal{N}(0, \sigma^{2}(t))$. 

We define the \textit{reference feature weights} as

$\bm{\beta}(t) \triangleq \left[ \bm{\beta}(t)_{1}, \bm{\beta}(t)_{2},...\bm{\beta}(t)_{K} \right]^{\top} \in \mathbb{R}^{K}$, where  \begin{equation}\label{eq:beta}
    \bm{\beta}(t)_k = \sum_{i=1}^{N} \beta_{k}(t) \cdot \bm{f}^{k}(t)_{i},
\end{equation} 
is the inner product of $ \frac{\partial (\bm{q}^{\top}(t) \cdot \bm{1})}{\partial \bm{f}^{k}(t)} = \beta_{k}(t) \cdot [1,...,1]^{\top}$ and $\bm{f}^{k}(t)$ that characterizes the total contribution of the $k$-th feature to the portfolio value at time $t$.

% The portfolio return at time slot $t$ is 
% $\bm{w}^{\top}(t) \cdot \bm{y}(t)= \bm{q}^{\top}(t) \cdot \bm{1}$. We use the gradient to measure the marginal contribution of features to portfolio return,
% \begin{equation}\label{beta_reference}
%     \frac{\partial (\bm{q}^{\top}(t) \cdot \bm{1})}{\partial \bm{f}^{k}(t)} = \beta_{k}(t) \cdot [1,...,1]^{\top}, ~~k = 1,...,K.
% \end{equation}
  
% Furthermore, we define $\bm{\beta}(t) \triangleq \left[ \bm{\beta}(t)_{1}, \bm{\beta}(t)_{2},...\bm{\beta}(t)_{K} \right]^{\top} \in \mathbb{R}^{K}$, where  \begin{equation}\label{eq:beta}
%     \bm{\beta}(t)_k = \sum_{i=1}^{N} \beta_{k}(t) \cdot \bm{f}^{k}(t)_{i},
% \end{equation} 
% is the inner product of $ \frac{\partial \bm{q}^{\top}(t) \cdot \bm{1}}{\partial \bm{f}^{k}(t)}$ and $\bm{f}^{k}(t)$ that characterizes the total contribution of the $k$-th feature to the portfolio value at time $t$. It serves as the reference for \textit{feature weights}.

% \begin{figure}[t]
% \centering
% \includegraphics[scale=0.44]{figs/intgradient.png}
% \caption{Overview of feature weight in DRL.}
% \label{trading}
% \end{figure}

\subsection{Feature Weights for DRL Trading Agent}

For a DRL agent in portfolio management task, at the beginning of a trading slot $t$, it takes  the feature vectors and co-variance matrix as input. Then it outputs an action vector, which is the portfolio weight vector $\bm{w}(t)$. We also represent it as a linear regression model,
\begin{equation}
    \bm{q}(t)
    = c_0(t)\cdot [1,...,1]^{\top} + c_1(t)\cdot\bm{f}^1(t) + ... +  c_K(t)\cdot\bm{f}^K(t) + \bm{\epsilon}(t).
\end{equation}

As Fig. \ref{int_grad} shows, for the decision-making process of a DRL agent, we define the feature weights for the $k$-th feature as

%\footnote{Assuming all the coefficients are capped by a positive constant, $|c_{k}(t)| < C $ , where $C \in \mathbb{R}^{+}$. Then the derivative and expectation is interchangeable.}

% \textbf{Saliency Map of DRL}\\
% % explain the figure 1, overview, workflow
% % 基础概念如何传达出去, 表达
\begin{equation}
\begin{split}
    &\bm{M}^{\pi}(t)_k \triangleq \sum_{i=1}^{N} IG(\bm{f}^{k}(t))_{i}  \\
    &\approx \sum_{i=1}^{N} \bm{f}^{k}(t)_{i} \cdot  \sum_{l=0}^{\infty} \gamma^{l}\cdot \frac{\partial\mathbb{E}\left[ \bm{w}^{\top}(t+l)\cdot\bm{y}(t+l) ) |\bm{s}^{k,i}(t),\bm{w}(t)\right]}{\partial \bm{f}^{k}(t)_{i}} \\
    &= \sum_{i=1}^{N} \bm{f}^{k}(t)_{i} \cdot  \sum_{l=0}^{\infty} \gamma^{l}\cdot \mathbb{E}\left[ c_{k}(t+l) \frac{\partial\bm{f}^{k}(t+l)_{i}}{\partial\bm{f}^{k}(t)_{i}}  |\bm{s}^{k,i}(t),\bm{w}(t)\right],
    %                   &= \sum_{i=1}^{N} (\bm{s}(t)_{k,i} - \bm{s}_{k}'(t)_{k,i})  \int_{z = 0}^{1} \frac{\partial Q^{\pi}(\bm{s}_{k}'(t) + z (\bm{s}(t) - \bm{s}_{k}'(t)), \bm{a}(t))}{\partial \bm{s}(t)_{k,i}} d z\\
    % &= \sum_{i=1}^{N} \bm{f}^{k}(t-1)_{i} \cdot  \frac{\partial Q^{\pi}(\bm{s}_{j}'(t) + z^{k,i} (\bm{s}(t) - \bm{s}_{k}'(t)), \bm{a}(t))}{\partial \bm{f}^{k}(t-1)_{i}} \cdot 1,
\end{split}
\end{equation}
where the last equality holds due to the fact that $\frac{\partial \bm{w}^{\top}(t+l) \cdot \bm{y}(t+l)}{\partial \bm{f}^{k}(t)_{i}}$ is continuous and $\bm{w}^{\top}(t+l) \cdot \bm{y}(t+l)$ is bounded for any $t$ \cite{enwiki:1037463814, enwiki:1036027918}.

Assuming the time dependency of features on stocks follows the power law, i.e.,
$\frac{\partial\bm{f}^{k}(t+l)_{i}}{\partial\bm{f}^{k}(t)_{i}}=l^{-\alpha}$, where $\alpha \in \mathbb{R}^{+}$, for $l \ge 1$, then the feature weights are
\begin{equation}
\begin{split}
    &\bm{M}^{\pi}(t)_k\\
    &= \sum_{i=1}^{N} \bm{f}^{k}(t)_{i} \cdot  \sum_{l=0}^{\infty} \gamma^{l}\cdot \mathbb{E}\left[ c_{k}(t+l) \frac{\partial\bm{f}^{k}(t+l)_{i}}{\partial\bm{f}^{k}(t)_{i}}  |\bm{s}^{k,i}(t),\bm{w}(t)\right] \\
    &= \sum_{i=1}^{N} \bm{f}^{k}(t)_{i} \cdot  \\
    &\{\mathbb{E}\left[ c_{k}(t) |\bm{s}^{k,i}(t),\bm{w}(t)\right]+\sum_{l=1}^{\infty} \gamma^{l}\cdot \mathbb{E}\left[ c_{k}(t+l) \cdot l^{-\alpha}  |\bm{s}^{k,i}(t),\bm{w}(t)\right]\} \\
    &=\sum_{i=1}^{N} \bm{f}^{k}(t)_{i} \cdot  \mathbb{E}\left[c_{k}(t) + \sum_{l=1}^{\infty} \gamma^{l} \cdot l^{-\alpha} \cdot c_{k}(t+l)  |\bm{s}^{k,i}(t),\bm{w}(t)\right].
\end{split}
\end{equation}
% where $z^{k,i} \in [0,1]$, $\frac{\partial Q^{\pi}(\bm{s}_{k}'(t) + z^{k,i} (\bm{s}(t) - \bm{s}_{k}'(t)), \bm{a}(t))}{\partial \bm{f}^{k}(t-1)_{i}}$ is the marginal contribution of Q value to the $k$-th feature at $i$-th stock.  $\bm{M}^{\pi}(t)_k$ is the inner product of marginal contribution and feature value at end of time slot $t$ under policy $\pi$.
% Comparing with $\beta_k(t)$, the major difference is
% \begin{itemize}[leftmargin=*]
%     \item  $\beta_k(t)$ in (\ref{beta_reference}) measures the marginal portfolio return at the $k$-th feature at $i$-th stock.
%     \item $\mathbb{E}\left[ c_{k}(t) |\bm{s}^{k,i}(t),\bm{a}(t)\right]$ measures the expected marginal portfolio return at the $k$-th feature at $i$-th stock.  
% \end{itemize}
% The difference also indicates the major advantage of the DRL agent, it makes decision with a long-term goal. 
Notice that $\bm{M}^{\pi}(t)_k$ has a similar form as $\bm{\beta}(t)_k$ in (\ref{eq:beta}). The $\beta_{k}(t)$ are replaced by $\mathbb{E}\left[c_{k}(t) + \sum_{l=1}^{\infty} \gamma^{l}\cdot l^{-\alpha} \cdot c_{k}(t+l)  |\bm{s}^{k,i}(t),\bm{w}(t)\right]$ in the context of a DRL agent. This better characterizes the superiority of the DRL agents to maximize future rewards.

% \begin{figure*}[ht]
% \centering
% \includegraphics[width=0.8\textwidth]{figs/modeldiagram.png}
% \caption{Portfolio management with forward-pass machine learning models and deep reinforcement learning models.}
% \label{portfolioPerformance}
% \end{figure*}

%%%%%%%%%%%%%%%%%%%%%%%%%%%%%%%%%%%%%%%%%%%%%%%%%%%%%%%
\subsection{Quantitative Comparison}

Our empirical approach provides explanations by quantitatively comparing the feature weights to the reference feature weights.\\
\textbf{Conventional Machine Learning Methods with Forward-Pass }\\
A conventional machine learning  method with a forward-pass has three steps: 1) Predict stock returns with machine learning methods using features. 2) Find optimal portfolio weights under predicted stock returns. 3) Build a regression model between portfolio return and features. 
%  \widehat{\bm{w}}^{*}(t) &= \frac{\widehat{\Sigma}^{-1}\cdot \widehat{\bm{\mu}}(t)}{\bm{1}^T\cdot \widehat{\Sigma}^{-1} \cdot \widehat{\bm{\mu}}(t)},\\
\begin{equation}
\begin{split}
    \widehat{\bm{y}}(t) &=  g(\bm{f}^1(t),...,\bm{f}^K(t)), \\
    \bm{q}^{*}(t) &= \bm{w}^{*}(t) \odot \bm{y}(t),\\ 
    \bm{q}^{*}(t) &= b_0(t)\cdot \bm{1} + b_1(t)\cdot\bm{f}^1(t) + ... +  b_K(t)\cdot\bm{f}^K(t) + \bm{\epsilon}(t),\\
    t &= 1,...,T,
\end{split}
\end{equation}
where  $g(\cdot)$ is the machine learning regression model. $b_k(t)$ is the gradient of the portfolio return to the $k$-th feature at time slot $t$, $i = 1,...,K$. $\bm{y}(t)$ is the true price relative vector at time $t$. $\widehat{\bm{y}}(t)$ is the predicted price relative vector at time $t$. $\bm{w}^{*}(t)$ is the optimal portfolio weight vector defined in (\ref{eq:opt_problem}), where we set the risk aversion parameter to 0.5. Likewise, we define the feature weights $\bm{b}(t)_{k}$ by
\begin{equation}
    \bm{b}(t)_{k} = \sum_{i=1}^{N} b_{k}(t) \cdot \bm{f}^{k}(t)_{i},
\end{equation} 
which is similar to how we define $\bm{\beta}(t)_{k}$ and $\bm{M}^{\pi}(t)_k$. \\
\textbf{Linear Correlations}\\
Both the machine learning methods and DRL agents take profits from their prediction power. We quantify the prediction power by calculating the linear correlations $\rho(\cdot)$ between the feature weights of a DRL agent and the reference feature weights and similarly for machine learning methods. 

Furthermore, the machine learning methods and DRL agents are different when predicting future. The machine learning methods rely on single-step prediction to find portfolio weights. However, the DRL agents find portfolio weights with a long-term goal. Then, we compare two cases, single-step prediction and multi-step prediction.  

For each time step, we compare a method's feature weights with $\bm{\beta}(t)$ to measure the single-step prediction. For multi-step prediction, we compare wih a smoothed vector,
\begin{equation}\label{eq:average_beta}
    \bm{\beta}^{W}(t) = \frac{\sum_{j = 0}^{W-1} \bm{\beta}(t+j)}{W},
\end{equation}
where $W$ is the number of time steps of interest. It is the average reference feature weights over $W$ steps.

% \begin{equation}\label{eq:correlation_coefficient}
% \begin{split}
%      \rho(\bm{M}^{\pi}(t), \bm{\beta}(t)) &= \frac{\text{Cov}[\bm{M}^{\pi}(t), \bm{\beta}(t)]}{\sqrt{\text{Var}[\bm{M}^{\pi}(t)] \cdot \text{Var}[\bm{\beta}(t)]}},\\
%      \rho(\bm{b}(t), \bm{\beta}(t)) &= \frac{\text{Cov}[\bm{b}(t), \bm{\beta}(t)]}{\sqrt{\text{Var}[\bm{b}(t)] \cdot \text{Var}[\bm{\beta}(t)]}}.
% \end{split}
% \end{equation}

\begin{figure}[t]
\centering
\includegraphics[scale=0.7]{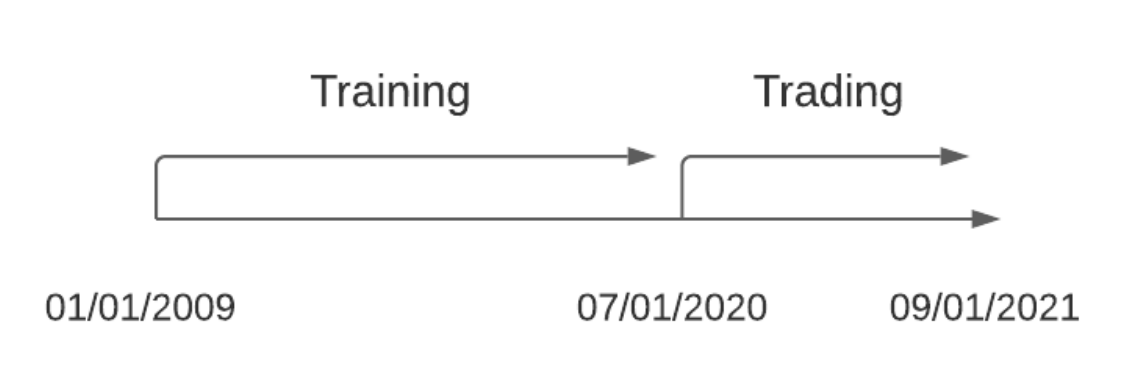}
\caption{Data split for the training and trading periods.}
\label{trading}
\end{figure}

\begin{figure*}[t]
\centering
\includegraphics[width=1\textwidth]{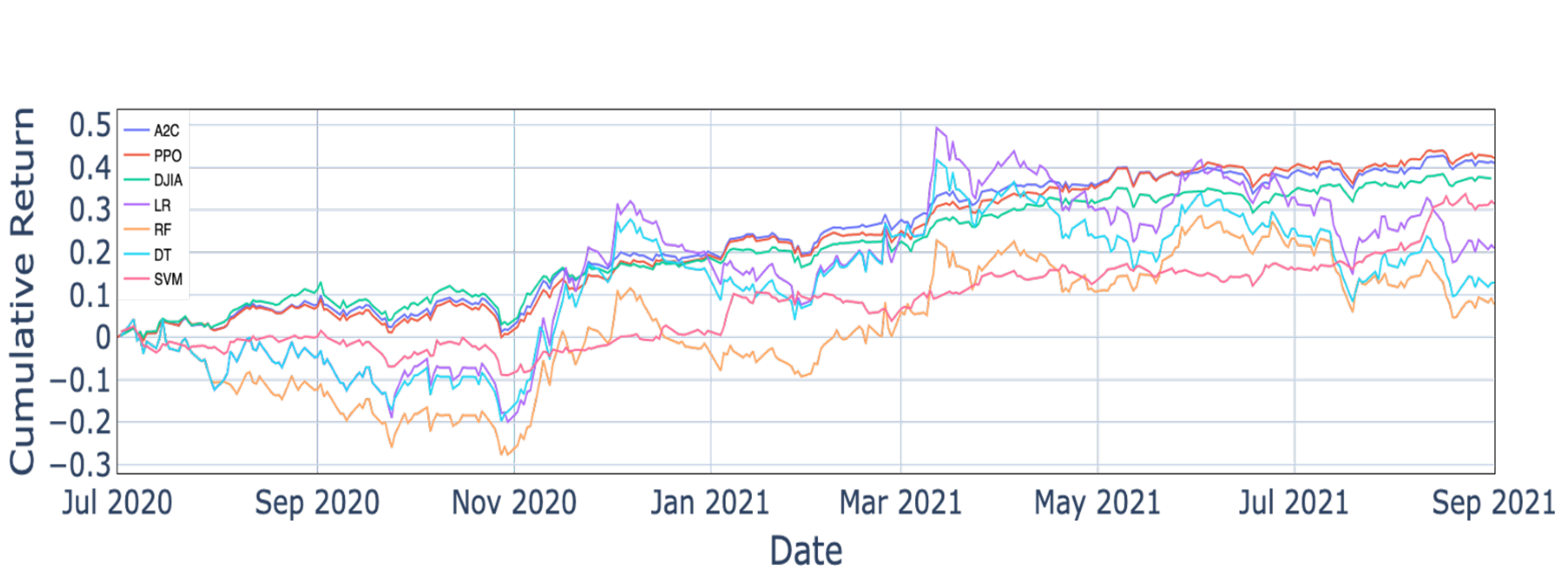}
\caption{The cumulative portfolio return curves of machine learning and DRL models (from 2020-07-01 to 2021-09-01).}
\label{fig:portfolioPerformance}
\end{figure*}

For $t=1,...,T$, we use the average values as metrics. For the machine learning methods, we measure the single-step and multi-step prediction power using
\begin{equation}
\label{eq:ML_prediction}
\begin{split}
    \overline{\rho}(\bm{b}, \bm{\beta}) &= \frac{\sum_{t=1}^{T} \rho(\bm{b}(t), \bm{\beta}(t))}{T},  \\
      \overline{\rho}(\bm{b}, \bm{\beta}^{W}) &= \frac{\sum_{t=1}^{T-W+1} \rho(\bm{b}(t), \bm{\beta}^{W}(t))}{T-W+1}.\\
\end{split}
\end{equation}
For the DRL-agents, we measure the single-step and multi-step prediction power using
\begin{equation}
\label{eq:DRL_prediction}
\begin{split}
     \overline{\rho}(\bm{M}, \bm{\beta}) &= \frac{\sum_{t=1}^{T} \rho(\bm{M}(t), \bm{\beta}(t))}{T},  \\
      \overline{\rho}(\bm{M}, \bm{\beta}^{W}) &= \frac{\sum_{t=1}^{T-W+1} \rho(\bm{M}(t), \bm{\beta}^{W}(t))}{T-W+1}.\\
\end{split}
\end{equation}

In (\ref{eq:ML_prediction}) and (\ref{eq:DRL_prediction}),
the first metric represents the average single-step prediction power during the whole trading period. The second metric then measures the average multi-step prediction power. 

These two metrics are important to explain the portfolio management task.
\begin{itemize}
    \item \textbf{Portfolio performance}: A closer relationship to the reference model indicates a higher prediction power and therefore a better portfolio performance. Both the single-step prediction and multi-step prediction power are expected to be positively correlated to the portfolio's performance.
    \item \textbf{The advantage of DRL agents}: The DRL agents make decisions with a long-term goal. Therefore the multi-step prediction power of DRL agents is expected to outperform their single-step prediction power.
    \item \textbf{The advantage of machine learning methods}: The portfolio management strategy with machine learning methods relies on single-step prediction power. Therefore, the single-step prediction power of machine learning methods is expected  to outperform their multi-step prediction power.
    \item \textbf{The comparison between DRL agents and machine learning methods}: The DRL agents are expected to outperform the machine learning methods in multi-step prediction power and fall behind in single-step prediction power.
\end{itemize}
 
% To investigate a DRL agent's performance in multi-step prediction, we compute a smoothed vector,
% \begin{equation}\label{eq:average_beta}
%     \overline{\bm{\beta}}(t) = \frac{\sum_{j = 0}^{W-1} \bm{\beta}(t+j)}{W},
% \end{equation}
% where $W$ is the number of time steps of interest. It is the average feature weight vector of reference model for the future $W$ steps. We can measure the model's multi-step prediction capability by plugging $\overline{\bm{\beta}}(t)$ in (23) back into (\ref{eq:correlation_coefficient}).

\section{Experimental Results}

In this section, we describe the data set, compared machine learning methods, trading performance and explanation analysis.

\begin{table*}[t]
\centering
\caption{Comparison of trading performance.}
\begin{tabular}{|l|c|c|c|c|c|c|c|}
\hline
\textbf{(2020/07/01-2021/09/01)}         & \textbf{PPO} & \textbf{A2C} & \textbf{DT} & \textbf{LR}   & \textbf{RF} & \textbf{SVM} & \textbf{DJIA}\\ \hline
\textbf{Annual Return}       & \textbf{35.0}\%                     & 34	\%                    & 10.8\%                          & 17.6\%                  & 6.5\%   & 26.2\%   & 31.2\%         \\ \hline
\textbf{Annual Volatility}   & \textbf{14.7}\%                     & 14.9	\%                    & 40.1\%                          & 42.4\%                    & 41.2	\%        & 16.2	\%  & 14.1	\%         \\ \hline
\textbf{Sharpe Ratio}        & \textbf{2.11}                      & 2.04                    & 0.45                        & 0.592                   & 0.36	  & 1.53 & 2.0 	                  \\ \hline
\textbf{Calmar Ratio}        & \textbf{4.23}                   & 4.30                      & 0.46                        & 0.76
                     & 0.21           & 2.33   & 3.5              \\ \hline
\textbf{Max Drawdown}        & -8.3\%                    & \textbf{-7.9}\%                   & -23.5\%                         & -23.2\%                  & -30.7	\%  & -11.3	\%   & -8.9	\%                \\ \hline
\textbf{Ave. Corr. Coeff. (single-step)}        & 0.024               & 0.030                   & \textbf{0.068}                         & 0.055                 & 0.052   & 0.034  & N/A           \\ \hline
\textbf{Ave. Corr. Coeff. (multi-step)}        & \textbf{0.09}                 & 0.078                 & -0.03                       & -0.03                 & -0.015   & -0.006  & N/A              \\ \hline
%\textbf{Daily Value at Risk} & -1.1\%                     & -1.7\%                    & -1.1\%                          & -1.5\%                   & -2.2\%  & -2.5\%                   \\ \hline
\end{tabular}
\label{table:performance_compare}
\end{table*}

\subsection{Stock Data and Feature Extraction}

We describe the stock data and the features.\\
\textbf{Stock data}. We use the {FinRL} library \cite{liu2020finrl} and the stock data of Dow Jones 30 constituent stocks, accessed at the beginning of our testing period, from 01/01/2009 to 09/01/2021.
The stock data is divided into two sets.
Training data set (from 01/01/2009 to 06/30/2020) is used to train the DRL agents and  machine learning models, while trading data set (from 07/01/2020 to 09/01/2021) is used for back-testing the trading performance.\\
\textbf{Features}. We use four technical indicators as features in our experiments.
\begin{itemize}[leftmargin=*]
    \item MACD: Moving Average Convergence Divergence.
    \item RSI: Relative Strength Index.
    \item CCI: The Commodity Channel Index.
    \item ADX: Average Directional Index.
\end{itemize}
All data and features are measured in a daily time granularity.

\begin{figure*}[t]
\centering
\includegraphics[width=1\textwidth]{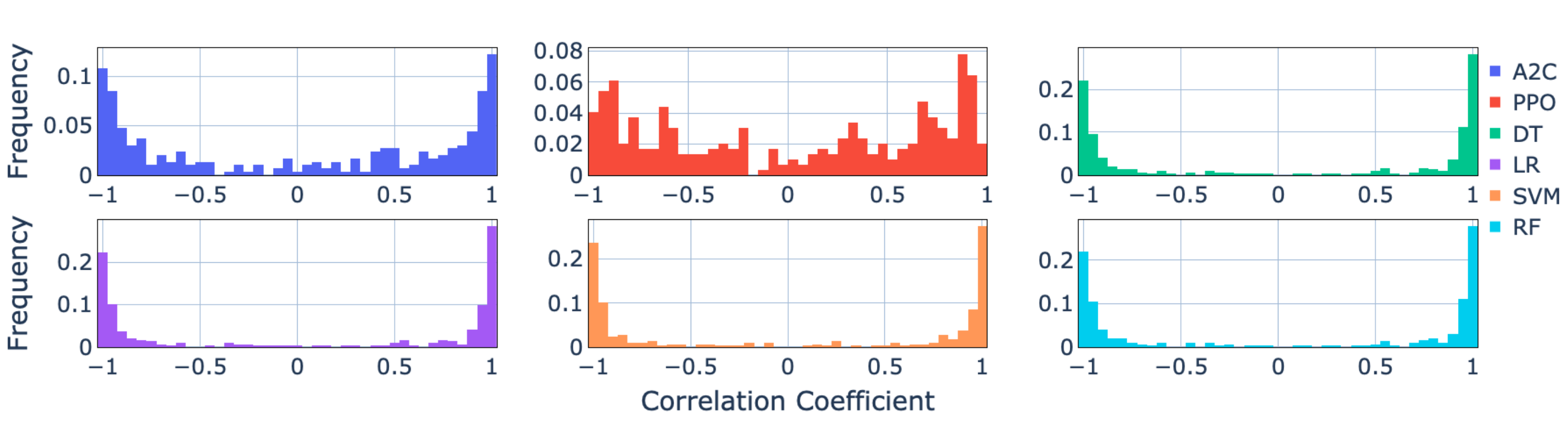}
\caption{The histogram of correlation coefficient (single-step) for Advantage Actor Critic (A2C), Proximal Policy Optimization (PPO), Decision Tree (DT), Linear Regression (LR), Support Vector Machine (SVM) and  Random Forest (RF).}
\label{fig:single}
\end{figure*}

\begin{figure*}[t]
\centering
\includegraphics[width=1\textwidth]{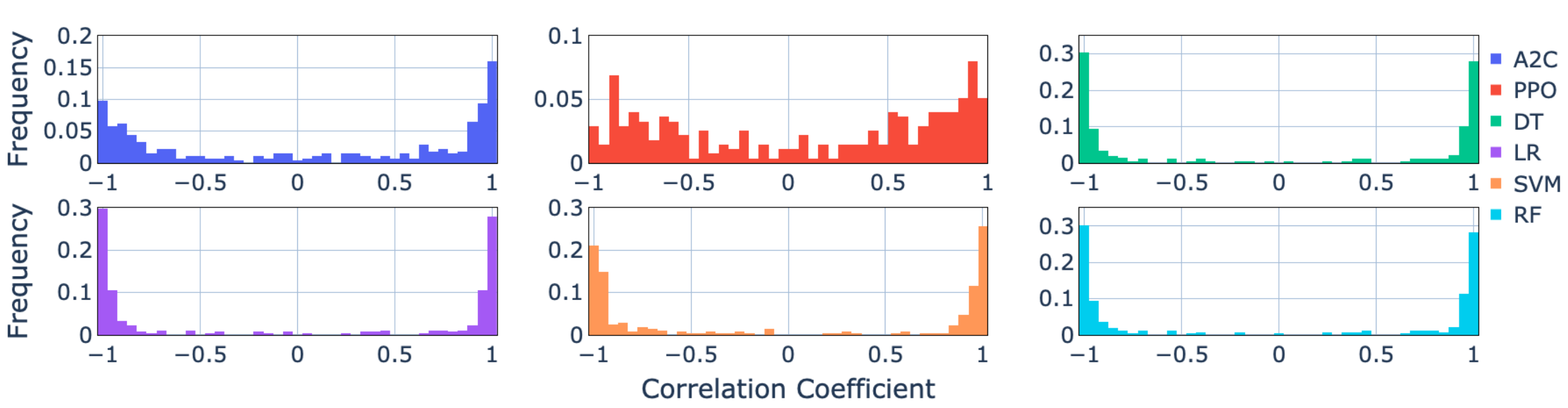}
\caption{The histogram of correlation coefficients (multiple-step) for Advantage Actor Critic (A2C), Proximal Policy Optimization (PPO), Decision Tree (DT), Linear Regression (LR), Support Vector Machine (SVM) and  Random Forest (RF).}
\label{fig:multi}
\end{figure*}

% \begin{table*}[t]
% \centering
% \caption{Upper tail test table for mean correlation coefficient (single-step) with significance level, $\alpha$.}
% \begin{tabular}{|l|c|c|c|c|c|c|}
% \hline
%         & \textbf{PPO} & \textbf{A2C} & \textbf{DT} & \textbf{LR}   & \textbf{RF} & \textbf{SVM} \\ \hline
% \textbf{Z-statistics}       & 0.41                    & 0.51	                    & \textbf{1.16}                         & 0.94                  & 0.90   & 0.58            \\ \hline
% \textbf{critical value ($\alpha$ = 10\%)}   & 1.282                  & 1.282	                    & 1.282                         & 1.282                    & 1.282	        & 1.282	           \\ \hline
% \textbf{critical value ($\alpha$ = 15\%)}   & 1.037                  & 1.037	                    & 1.037                         & 1.037                   & 1.037        & 1.037	           \\ \hline
% %\textbf{Daily Value at Risk} & -1.1\%                     & -1.7\%                    & -1.1\%                          & -1.5\%                   & -2.2\%  & -2.5\%                   \\ \hline
% \end{tabular}
% \label{table:stat-tests-single}
% \end{table*}

\begin{table*}[t]
\centering
\caption{Upper tail test table for mean correlation coefficient (single-step and multi-step) under null hypothesis: the mean correlation coefficient is of no difference than zero.}
\begin{tabular}{|l|c|c|c|c|}
\hline
        & $\textbf{Z-statistics (single-step)}$ 
        & $\textbf{Z-statistics (multi-step)}$
        \\ \hline
$\textbf{PPO}$    & 0.6   & $\textbf{2.16}^{***}$  \\ \hline
$\textbf{A2C}$   & 0.51                 	                          & $\textbf{1.58}^{**}$                 	                          \\ \hline
$\textbf{DT}$   & $\textbf{1.28}^{**}$                                   & -0.59                                                   \\ \hline
$\textbf{LR}$   & 1.03                 	                            & -0.55                	                                    \\ \hline
$\textbf{RF}$   & 0.98                  	                & -0.28                           	                 \\ \hline
$\textbf{SVM}$   & 0.64                 	                           & -0.11                 	                                        \\ \hline
%\textbf{Daily Value at Risk} & -1.1\%                     & -1.7\%                    & -1.1\%                          & -1.5\%                   & -2.2\%  & -2.5\%                   \\ \hline
\end{tabular}
\label{table:stat-tests-single}
\end{table*}

% \begin{table*}[t]
% \centering
% \caption{Upper tail test table for mean correlation coefficient (multi-step) with significance level, $\alpha$.}
% \begin{tabular}{|l|c|c|c|c|c|c|}
% \hline
%         & \textbf{PPO} & \textbf{A2C} & \textbf{DT} & \textbf{LR}   & \textbf{RF} & \textbf{SVM} \\ \hline
% \textbf{Z-statistics}       & 1.49                    & 1.28	                                            & -0.55                  & -0.51   & -0.26        & -0.1     \\ \hline
% \textbf{critical value ($\alpha$ = 10\%)}   & 1.282                  & 1.282	                    & 1.282                         & 1.282                    & 1.282	        & 1.282	           \\ \hline
% \textbf{critical value ($\alpha$ = 15\%)}   & 1.037                  & 1.037	                    & 1.037                         & 1.037                   & 1.037        & 1.037	           \\ \hline
% %\textbf{Daily Value at Risk} & -1.1\%                     & -1.7\%                    & -1.1\%                          & -1.5\%                   & -2.2\%  & -2.5\%                   \\ \hline
% \end{tabular}
% \label{table:stat-tests-multi}
% \end{table*}
\subsection{Compared Machine Learning Methods}

We describe the models we use in experiment.
We use four classical machine learning regression models \cite{scikit-learn}: Support Vector Machine (SVM), Decision Tree Regression (DT), Linear Regression (LR), Random Forest (RF) and two deep reinforcement learning models: A2C and PPO.

% \begin{figure}[t]
% \centering
% \includegraphics[scale=0.9]{figs/train_split.png}
% \caption{The split of train-test-trade.}
% \label{trading}
% \end{figure}

% \begin{figure}[t]
% \centering
% \includegraphics[scale=0.18]{figs/ROC_IC.png}
% \caption{The information coefficient statistics of ROC}
% \label{trading}
% \end{figure}

% \begin{figure}[t]
% \centering
% \includegraphics[scale=0.18]{figs/MFI_IC.png}
% \caption{The information coefficient statistics of MFI.}
% \label{trading}
% \end{figure}

% \begin{figure}[t]
% \centering
% \includegraphics[scale=0.17]{figs/Aroon_IC.png}
% \caption{The information coefficient statistics of Aroon}
% \label{trading}
% \end{figure}

% \begin{figure}[t]
% \centering
% \includegraphics[scale=0.17]{figs/RSI_IC.png}
% \caption{The information coefficient statistics of RSI}
% \label{trading}
% \end{figure}

% \begin{figure}[t]
% \centering
% \includegraphics[scale=0.2]{figs/boxplot.png}
% \caption{The non-linear effect between positive and negative returns}
% \label{trading}
% \end{figure}

% \begin{figure}[t]
% \centering
% \includegraphics[scale=0.17]{figs/pvaluePlot.png}
% \caption{The p-value of mean/t score of different window size}
% \label{trading}
% \end{figure}

\subsection{Performance Comparison}

% \begin{figure}[t]
% \centering
% \includegraphics[scale=0.14]{figs/algo_compare.png}
% \caption{The algorithms comparison}
% \label{trading}
% \end{figure}
We use several metrics to evaluate the trading performance.
\begin{itemize}[leftmargin=*]
    \item \textbf{Annual return}: the geometric average portfolio return each year.
    \item \textbf{Annual volatility}: The annual standard deviation of the portfolio return.
    \item \textbf{Maximum drawdown}: The maximum percentage loss during the trading period.
    \item \textbf{Sharpe ratio}: The annualized portfolio return in excess of the risk-free rate per unit of annualized volatility.
    \item \textbf{Calmar ratio}: The average portfolio return per unit of maximum drawdown.
    \item \textbf{Average Correlation Coefficient (single-step)}: It measures a model's single-step prediction capability.
    \item \textbf{Average Correlation Coefficient (multi-step)}: It measures a model's multi-step prediction capability.  We set $W = 20$ in (\ref{eq:average_beta}).
\end{itemize}
 As shown in by Fig. \ref{fig:portfolioPerformance} and Table \ref{table:performance_compare}, the DRL agent using PPO reached 35$\%$ for annual return and 2.11 for Sharpe ratio, which performed the best among all the others. The other DRL agent using A2C reached 34$\%$ for annual return and 2.04 for Sharpe ratio. Both of them performed better than the Dow Jones Industrial Average (DJIA), which reached 31.2$\%$ for annual return and 2.0 for Sharpe ratio. As for the machine learning methods, the support vector machine method reached the highest Sharpe ratio: 1.53 and the highest annual return: 26.2$\%$. None of the machine learning methods outperformed the Dow Jones Industrial Average (DJIA).
 
% \begin{itemize}[leftmargin=*]
%     \item Only PPO and A2C has the largest Sharpe ratio and Calmar ratio.
%     \item Linear regression outperforms other machine learning models in terms of Sharpe ratio and Calmar ratio.
% \end{itemize}
\begin{figure*}[t]
\centering
\includegraphics[width=0.9\textwidth]{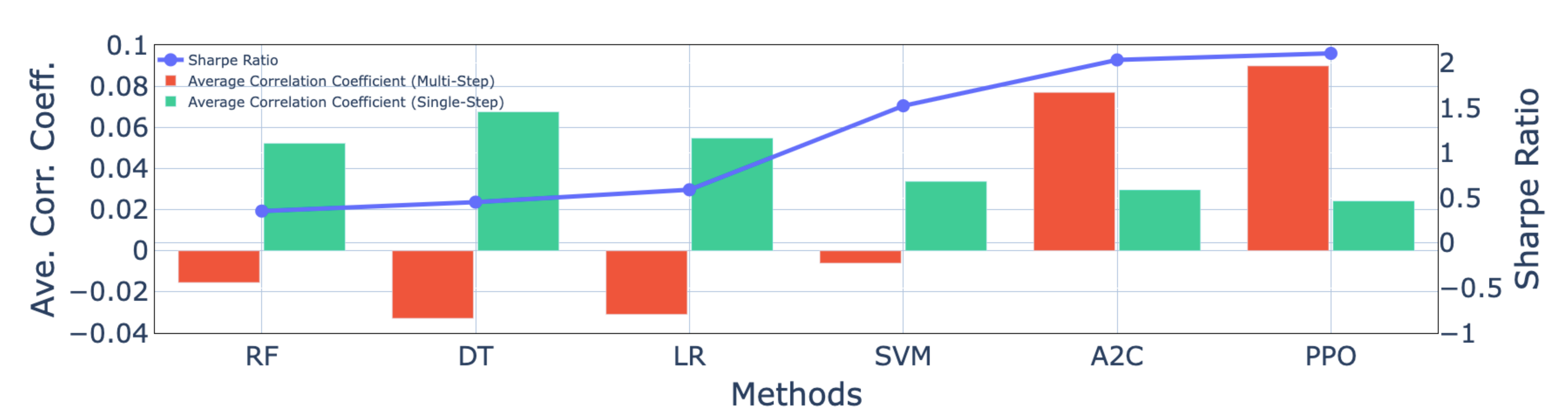}
\caption{The comparison of Sharpe ratio and average correlation coefficient.}
\label{fig:coefficient_compare}
\end{figure*}

\subsection{Explanation Analysis}

We calculate the histogram of correlation coefficients with 1770 samples for 295 trading days.  From Fig. \ref{fig:single} and Fig. \ref{fig:multi}, we visualize the the distribution of correlation coefficients. We derived the statistical tests as in 
Table \ref{table:stat-tests-single}, where  "\textbf{**}", "\textbf{***}" denote significance at the 10$\%$ and  5$\%$ level.  We find that
\begin{itemize}
    \item The distributions of correlation coefficients are different between the DRL agents and machine learning methods.
    \item The machine learning methods show greater significance in mean correlation coefficient (single-step) than DRL agents.
     \item The DRL agents show stonger significance in mean correlation coefficient (multi-step) than machine learning methods.
\end{itemize}

We show our method empirically explains the superiority of DRL agents for the portfolio management task. As Fig. \ref{fig:coefficient_compare} shows, the y-axis represents the average coefficients and Sharpe ratio for the whole trading data set, the x-axis represents the model.  
From Table 1 and Fig. \ref{fig:coefficient_compare}, we find that 
\begin{itemize}[leftmargin=*]
    \item The DRL agent using PPO has the highest Sharpe ratio:2.11 and highest average correlation coefficient (multi-step): 0.09  among all the others.
    \item The DRL agents' average correlation coefficients (multi-step) are significantly higher than their average correlation coefficients (single-step).
    \item The machine learning methods' average correlation coefficients (single-step) are significantly higher than their average correlation coefficients (multi-step).
    \item The DRL agents outperform the machine learning methods in  multi-step prediction power and fall behind in single-step prediction power.
    \item Overall, a higher mean correlation coefficient (multi-step) indicates a higher Sharpe ratio.
    % \item Higher average correlation coefficients (both one-step and multi-step) indicate  better performance.
    %  \item The average correlation coefficients (single-step) for machine learning models are higher than DRL agents. It indicates that machine learning models are better at single-step prediction. 
    % \item The average correlation coefficients (multi-step) for machine learning models are negative. It indicates that machine learning models are weak at multi-step prediction. 
    % \item The average correlation coefficient (multi-step) of DRL agents are higher than machine learning models, which indicates the DRL agent performs better at multi-step prediction than machine learning models. 
\end{itemize}

%%为什么用在金融上?怎么交易的周末操作的?这样更新为什么成功?input和output，目标函数和结构?
%不做科普,模型互补

%%%%%%%%%%%%%%%%%%%%%%%%%%%%%%%%%%%%%%%%%%%%%%%%%%%%%%%%%%%%%%

% \begin{figure}[t]
% \centering
% \includegraphics[scale=0.18]{figs/multi_box.png}
% \caption{ Box plot of multi-period score}
% \label{trading}
% \end{figure}

\section{Conclusion}

In this paper, we empirically explained the DRL agents' strategies for the portfolio management task. We used a \textit{linear model in hindsight} as the \textit{reference model}.  We found out the relationship between the reward (namely, the portfolio return) and the input (namely, the features) using integrated gradients. We measured the prediction power using correlation coefficients.

We used Dow Jones 30 constituent stocks from 01/01/2009 to 09/01/2021 and empirically showed that DRL agents outperformed the machine learning models in multi-step prediction. For future work, we will explore the explanation methods for other deep reinforcement learning algorithms and study on other financial applications including trading, hedging and risk management.

 %observe intelligent behaviors \cite{ITS2018},, and incorporate prediction schemes \cite{Time_Series}.


\bibliographystyle{ACM-Reference-Format}
\bibliography{references}

\appendix

\end{document}